	\documentclass[journal,onecolumn]{IEEEtran}
 \UseRawInputEncoding 
	\usepackage{graphicx}          
	\usepackage{tikz} 
	\usetikzlibrary{shapes.geometric}
	\usetikzlibrary{arrows.meta,arrows,automata,positioning, shapes, backgrounds}
	\usetikzlibrary{calc}
	\usetikzlibrary{circuits.ee.IEC}
	
	\usepackage{textcomp}
	\usepackage{textcomp}

	\usepackage{algorithm}
	\usepackage[noend]{algpseudocode}

	\usepackage{amssymb}
	\usepackage{amsmath}
	\usepackage{bigints}
	\usepackage{enumerate}
	\usepackage{xcolor}

	\newtheorem{remark}{Remark} 

	\newtheorem{lemma}{Lemma} 
	\newtheorem{example}{Example} 
	\newtheorem{theorem}{Theorem} 
	\DeclareMathAlphabet{\mathpzc}{OT1}{pzc}{m}{it}
\begin{document}
	\title{Fisher Information of a Family of Generalized Normal Distributions}
	\author{Precious~Ugo~Abara and Sandra~Hirche
	\thanks{Precious Ugo~Abara and Sandra~Hirche are with Technical University of Munich, Germany,
		Department of Electrical and Computer Engineering,
		Chair of Information-oriented Control (ITR) e-mail: \{ugoabara, hirche\}@tum.de.}
	}

	\maketitle
	\begin{abstract}
		In this brief note we compute the Fisher information of a family of generalized normal distributions.
		Fisher information is usually defined for regular distributions, i.e. continuously differentiable (log) density functions whose support does not depend on the family parameter $\theta$. Although the uniform distribution in $[-\theta, + \theta]$ does not satisfy the regularity requirements, as a special case of our result, we will obtain the Fisher information for this family.
	\end{abstract}
	
	
	
	\section{Introduction}
	\IEEEPARstart{A}{} natural question is that arises in the context of parameter estimation is: how much information can a sample of data provide about the unknown parameter? This is a widely studied problem in literature \cite{kullback1951information, cover2012elements}. One information metric is the Fisher information \cite{lehmann2006theory}. Given a family of density functions $f(x, \theta)$ function of a parameter $\theta$, it always hold
	\begin{equation}
		\label{eq: density = 1}
		\int f(x, \theta) {\rm d} x = 1.
	\end{equation}
	Under certain regularity conditions, if $\theta$ is the true parameter, it can be shown that of \eqref{eq: density = 1} translates into
	\begin{equation*}
		\mathbb{E} \left[ \frac{\partial}{\partial \theta} \operatorname{log} f(x, \theta) \Big\vert \theta \right] = 0.
	\end{equation*}
	The partial derivate ${\partial}/{\partial \theta}$ of the natural logarithm of the likelihood $f(x, \theta)$ is referred to as \text{score}. The Fischer information $\mathcal{I}(\theta)$is defined as the variance of the score i.e. 
	\begin{equation*}
		\mathcal{I}(\theta) = \mathbb{E} \left[ \left(\frac{\partial}{\partial \theta} \operatorname{log} f(x, \theta)\right)^2 \Big\vert \theta \right].
	\end{equation*}
	If the second derivative with respect to $\theta$ of $\operatorname{log} f(x, \theta)$ exists for all $x$ and $\theta$ and the second derivative with respect to $\theta$ of the left side of \eqref{eq: density = 1} can be obtained by differentiating twice under the integral sign then the Fisher information may also be written as \cite{lehmann2006theory}
	\begin{equation}
		\label{eq: fisher 2}
		\mathcal{I}(\theta) = - \mathbb{E} \left[\frac{\partial^2}{\partial \theta^2} \operatorname{log} f(x, \theta) \Big\vert \theta \right].
	\end{equation}
	We note that the condition of interchange of the derivative and integration excludes the uniform distribution $f(x, \theta) = \frac{1}{\theta}$, $\theta \in [a, b]$. In order words, the uniform distribution does not meet the regularity conditions.
	\section{Preliminaries}
\begin{figure}[tb!]
		\begin{center}
			{\includegraphics[width=0.5\textwidth]{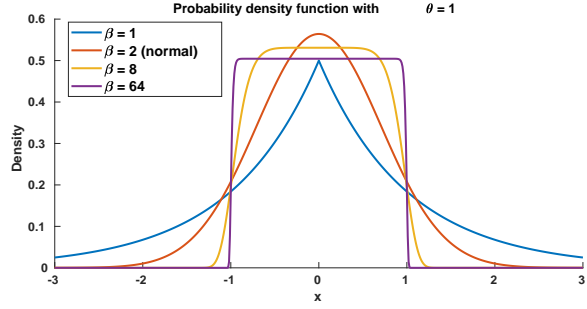}}
			\caption{Generalized normal distribution for $\theta = 1$ and different shape parameter $\beta$.} 
			\label{fig: comparison shape parameter}
		\end{center}
	\end{figure}

	In this section we give results for a random variable $X$ with a generalized normal distribution with zero mean i.e.
	\begin{equation}
		\label{eq: zero mean generalized normal distribution}
		f(x, \theta) = \frac{\beta}{2 \theta \Gamma(1/\beta)} \operatorname{exp}\left(-\frac{\vert x \vert^\beta}{\theta^\beta} \right)
	\end{equation}
	where $\Gamma$ denotes the gamma function, $\theta>0$ is called the scale, $\beta > 0$ is the shape. For a complex variable $z$, $\Gamma$ is defined as
	\begin{equation*}
		\Gamma (z) = \begin{cases}
		(z  - 1)!, & \text{if } z \text{ is positive integer},\\
		\int_{0}^{\infty} x^{z-1}\operatorname{exp}(-x) {\rm d}x, & \text{if }\mathbb{R}(z) >0.
		\end{cases}
	\end{equation*}
	Equation \eqref{eq: zero mean generalized normal distribution} is a parametric family of symmetric distributions. It includes the zero-mean Laplace distribution when $\beta = 1$, the zero-mean normal distribution when $\beta = 2$. Furthermore, it converges point-wise to the uniform density for $\beta \to \infty$. See Fig.~\ref{fig: comparison shape parameter}.
	\begin{lemma}[Moments \cite{nadarajah2005generalized}]
	Let $X$ be zero mean generalized Gaussian distribution of shape $\beta$ and scaling parameter $\theta$ as in \eqref{eq: zero mean generalized normal distribution}. The moments of $X$ exist and are finite for any $k$ greater than $−1$. For any non-negative integer $k$, the plain central moments are		
	\begin{equation*}
		\mathbb{E} \left[X ^ k\right] = \begin{cases}
		\theta^k \frac{\Gamma\left(\frac{k + 1}{\beta}\right)}{\Gamma\left(\frac{1}{\beta}\right)}, & \text{if } k \text{ is even},\\
		0, & \text{otherwise}.
		\end{cases}
	\end{equation*}
	\end{lemma}
	\begin{lemma}[$\Gamma$ at some rational arguments]
		Given integers $n>0$ and $p>0$ it holds that
		\begin{equation}
			\label{eq: gamma rational argument}
			\Gamma\left(n + \frac{1}{p}\right) = \Gamma\left(\frac{1}{p}\right)\frac{(pn - (p - 1))!^{(p)}}{p^n}
		\end{equation}
		where $n!^{(p)}$ denotes the $p$-th multi-factorial of $n$.
	\end{lemma}
	\begin{IEEEproof}
		Using the functional equation $\Gamma(\alpha + 1) = \alpha \Gamma(\alpha)$ for $\alpha>0$, and by its successive application, it is easy to obtain
		\begin{equation*}
			\Gamma\left(n + \frac{1}{p}\right) = \Gamma\left(\frac{1}{p}\right)\frac{(pn - p + 1)\left(p(n-1) - p +1\right) \cdots \left(p(n-n+1) -p + 1\right))}{p^n}.
		\end{equation*}
		This concludes the proof.
	\end{IEEEproof}
	\section{Result}
	\begin{theorem}
		Suppose $X$ is a random variable with a zero-mean generalized normal distribution given by \eqref{eq: zero mean generalized normal distribution}, and an even shape $\beta$, i.e. $\beta = 2n$, with $n$ a positive integer. Then
		\begin{equation*}
			\mathcal{I}(\theta) = \frac{\beta}{\theta^2}.
		\end{equation*}
	\end{theorem}
	\begin{IEEEproof}
		Since \eqref{eq: zero mean generalized normal distribution} is twice differentiable we can use \eqref{eq: fisher 2} to compute the Fisher information. The log-likelihood is given by
		\begin{equation*}
			\operatorname{log} f(x, \theta) = \operatorname{log}\frac{\beta}{2} - \operatorname{log}\theta - \operatorname{log}\Gamma\left(\frac{1}{\beta}\right) - \frac{\vert x\vert^\beta}{\theta^\beta}.
		\end{equation*}
		Its first partial derivative with respect to the parameter $\theta$ is
		\begin{equation*}
			\frac{\partial}{\partial \theta} \operatorname{log} f(x, \theta) = -\frac{1}{\theta} + \frac{\beta \theta^{\beta - 1}}{\theta^{2\beta}}\vert x\vert^\beta = -\frac{1}{\theta} + \frac{\beta\vert x\vert^\beta}{\theta^{\beta + 1}}.
		\end{equation*}
		The second derivative then follows as
		\begin{equation}
			\label{eq: second derivative}
			 \frac{\partial^2}{\partial \theta^2} \operatorname{log} f(x, \theta) = \frac{1}{\theta^2} - \frac{\beta(\beta + 1)\theta^{\beta}}{\theta^{2\beta + 2}}\vert x\vert^\beta = \frac{1}{\theta^2} - \frac{\beta(\beta + 1)\vert x\vert^\beta}{\theta^{\beta + 2}}.
		\end{equation}
		Using \eqref{eq: fisher 2} and \eqref{eq: second derivative} it then follows 
		\begin{equation}
			\label{eq: x fisher expectation}
			\mathcal{I}(\theta) = - \mathbb{E} \left[\frac{\partial^2}{\partial \theta^2} \operatorname{log} f(x, \theta) \Big\vert \theta \right] = - \frac{1}{\theta^2} + \frac{\beta(\beta + 1) \mathbb{E} \left[\vert x\vert^\beta \vert \theta \right]}{\theta^{\beta + 2}}.
		\end{equation}
		Since $\beta$ is assumed to be a positive even integer it holds that 
		\begin{equation}
			\label{eq: expection x^beta}
			\mathbb{E} \left[\vert x\vert^\beta \vert \theta \right] = \mathbb{E} \left[x^\beta \vert \theta \right] = \theta^\beta \frac{\Gamma\left(\frac{\beta + 1}{\beta}\right)}{\Gamma\left(\frac{1}{\beta}\right)} \overset{\eqref{eq: gamma rational argument}}{=} \frac{\theta^\beta}{\beta} \frac{\Gamma\left(\frac{1}{\beta}\right)}{\Gamma\left(\frac{1}{\beta}\right)} = \frac{\theta^\beta}{\beta}.
		\end{equation}
		where we used \eqref{eq: gamma rational argument} with $p = \beta$, $n = 1$ and noticing that $1!^{(\beta)} = 1$. Substituting \eqref{eq: expection x^beta} in \eqref{eq: x fisher expectation}
		\begin{equation*}
			\mathcal{I}(\theta) = - \frac{1}{\theta^2} + \frac{\beta + 1}{\theta^{2}} = \frac{\beta}{\theta^2}.
		\end{equation*}
	\end{IEEEproof}
	\begin{example}
		A zero-mean Gaussian distribution is obtained from \eqref{eq: zero mean generalized normal distribution} with $\beta = 2$. It is easy to see that a family of Gaussian distributions with unknown \textit{standard deviation} has Fisher information $\mathcal{I}(\theta) = \frac{2}{\theta^2}$.
	\end{example}
	\begin{remark}
		When $\beta \to \infty$ the distribution \eqref{eq: zero mean generalized normal distribution} converges to a uniform density on $(-\theta, +\theta)$. The Fisher information is thus $\mathcal{I}(\theta) = \infty$, for $\beta \to \infty$.
	\end{remark}
	\bibliographystyle{IEEEtran}
	\bibliography{References/refs.bib}
\end{document}